\pdfoutput=1
\documentclass[a4paper,american,aps,citeautoscript,floatfix,pdftex,pra,preprintnumbers,showpacs,superscriptaddress,twoside,%
longbibliography,%
reprint,twocolumn,final
]{revtex4-1}
\usepackage{amsfonts,amsmath,amssymb}
\usepackage{graphicx}
\usepackage[T1]{fontenc}
\usepackage{lmodern}
\usepackage{natbib}
\usepackage[utf8]{inputenc}
\usepackage{microtype}
\usepackage[textsize=footnotesize,obeyFinal]{todonotes}
\usepackage{hyperref}
\usepackage{upgreek}

\newlength{\figwidth}
\newcommand{\iop}{\affiliation{Institute of Physics, Faculty of Physics, Astronomy and Informatics, Nicolaus 
Copernicus University, Grudziadzka 5, 87-100 Torun, Poland}}%
\newcommand{\vsemail}{\email{vijay.singh@fizyka.umk.pl}}%
\begin{document}
\title{Theoretical investigation of a two-stage buffer gas cooled beam source}%
\author{Vijay Singh}\vsemail\iop%
\date{\today}%

\begin{abstract}\noindent%
A novel two-stage helium buffer gas cooled beam source is introduced. The properties of the molecular beams produced from this source are investigated theoretically using the CaF as a test molecule. The gas-phase molecules are first produced inside a 3~K helium buffer gas cell by laser ablation and subsequently cooled down to 3~K by collisions with buffer gas atoms. The precooled molecules are then extracted into the 0.5~K helium buffer gas cell where they are cooled further down to 0.5~K by collisions with cold helium atoms. Finally, the cold molecules are extracted out into the high vacuum through the 0.5~K cell exit aperture and form a molecular beam. The mean forward velocity and the beam flux are calculated to be 45~m/s and 8$\times$10$^{12}$ molecules per pulse respectively when both cells are operated in the so-called hydrodynamic entrainment regime. Using this flux and Maxwell-Boltzmann probability density function at 0.5~K, the number of the molecules moving with speeds $\leq$~5~m/s is calculated to be 8$\times$10$^{9}$. These slow and intense beams of the cold molecules are beneficial for efficient magneto-optical trapping of the molecules, investigating sympathetic cooling of the molecules with ultracold atoms, and performing ultrahigh precision molecular spectroscopy. 
\end{abstract}
\maketitle%

\section{Introduction}
\label{sec:introdution}
The collisions of the cold buffer gas atoms with warm molecules cool their translational and rotational temperatures~
\cite{Weinstein:Nature395:148,Hutzler:CR112:4803,Singh:PRL108:203201}. Thus, cryogenic helium buffer gas-cooled beams are prerequisite for the laser cooling of the molecules~\cite{Anderegg:PRL119:103201,Shuman:Nature467:820}. Further, the number of photon absorption and emission cycles required for the laser cooling of atoms and molecules decrease with their forward velocity and hence the laser cooling efficiency of the molecules increases with decrease in their forward velocity~
\cite{Anderegg:PRL119:103201,Kozyryev:PRL118:173201,Williams:PRL120:163201,Shuman:Nature467:820,Zhelyazkova:PRA89:053416}. 
In addition to this, the slower beams are also beneficial for efficient deceleration of the atoms and molecules by moving trap decelerators~
\cite{Ofir:NJP13:103030,McArd:MTZD:inprep,Perez:PRL110:133003}. This is because the reduced initial velocity facilitates in producing deeper traps. This results in increased velocity acceptance and hence the overall efficiency of the decelerator. Furthermore, the intense and slow beams of buffer gas-cooled molecules increase the product~$\sqrt{N}\tau$, where $N$ is the total number of molecules that participate in the experiments and $\tau$ is the measurement time. Consequently, they enhance the statistical sensitivity of the ultrahigh precision measurements~
\cite{Hudson:Nature473:493,ACME:Science343:269,QUACK:CPL132:147,Tokunaga:NJP19:053006,Kozyryev:PRL119:133002}. 

To extract the maximum number of molecules from the buffer gas cell into the molecular beam, the buffer gas cell is operated in the so-called hydrodynamic entrainment regime~
\cite{Hutzler:CR112:4803,Patterson:JCP126:154307}. This regime occurs at higher buffer gas densities such that the collisions of the buffer gas atoms at the cell exit boost the mean velocity of the molecules as high as the mean forward velocity ($v_\text{f} = \sqrt{\frac{2k_\text{B}T}{m}}$) of the buffer gas atoms. Where $k_\text{B}$ is Boltzmann constant, and $T$ and $m$ are the temperature and mass of the buffer gas atoms respectively. This boosting can however be suppressed partially by collisions of the molecules with buffer gas atoms inside a low buffer gas density slowing cell attached to the main buffer gas cell~\cite{Anderegg:PRL119:103201,Lu:PCCP13:18986}. But due to the accelerated diffusion losses inside the low-density buffer gas cell, the fraction of the molecules extracted from the cell into the molecular beam is reduced by several orders of magnitude~\cite{Patterson:JCP126:154307,Lu:PCCP13:18986}.

This paper introduces a new approach for reducing the forward velocity of the buffer gas-cooled beams without compromising the flux of molecules. Our approach is based on operating the buffer gas cell at the lowest temperature for minimizing the boosting of the molecules. At 0.5~K, the calculated mean forward velocity of the helium atoms is 45~m/s and hence the mean forward velocity of the helium buffer gas-cooled molecules can be reduced to 45~m/s by lowering the buffer gas cell temperature down to 0.5~K, even though the cell is operated in hydrodynamic entrainment regime. Furthermore, as discussed in~\autoref{sec:PDF}, the Maxwell-Boltzmann probability density function (PDF) of slower buffer gas atoms increases rapidly with decrease in temperature and hence the flux of the buffer gas-cooled molecules moving with speeds $\leq$~5~m/s is enhanced by a factor of 25 when the buffer gas cell temperature is lowered from 4~K to 0.5~K. This enhancement in the flux of the slower molecules is beneficial for efficient loading of the molecules into a magneto-optical trap~\cite{Anderegg:PRL119:103201,Williams:PRL120:163201}, investigating sympathetic cooling of buffer gas-cooled molecules with ultracold atoms~\cite{Lim:PRA92:053419}, and performing ultrahigh precision molecular spectroscopy~
\cite{Hudson:Nature473:493,ACME:Science343:269,QUACK:CPL132:147,Tokunaga:NJP19:053006,Kozyryev:PRL119:133002}.

Using the $^4$He as a buffer gas, a density of 10$^{15}$~cm$^{-3}$ can be achieved at 0.5~K cell temperature~\cite{Pobell:17675,Bretislav:JCS94:1783}. While this density is sufficient for the buffer gas cooling of the atoms and molecules, the heat load on the buffer gas cell from the vaporization techniques such as laser ablation usually exceeds the cooling power of the evaporation refrigerators at 0.5~K~\cite{Anderegg:PRL119:103201,Singh:PRL108:203201}. To implement this new approach, avoiding the heat load on the 0.5~K stage of the helium buffer gas cell from the laser ablation is therefore a prerequisite. This can be achieved by using a novel two-stage helium buffer gas cooling technique introduced here.

The first stage helium buffer gas cell is attached to the second stage of the cryogenic refrigerator and cooled below 3~K~\cite{Singh:PRL108:203201,Singh:PRA97:032704}. On the other hand, the second stage buffer gas cell is attached to the $^3$He bath and cooled down to 0.5~K by evaporative cooling~\cite{Anderegg:PRL119:103201,Singh:PRL108:203201}. Due to the significantly higher vapor pressure of the $^3$He compared to $^4$He at low temperatures~\cite{Pobell:17675,HUANG:CR46:833}, a higher cooling power can be achieved by pumping on a $^3$He bath at 0.5~K~\footnote{From the relation, $dP/dT = LP/RT^2$, the cooling power of an evaporation refrigerator is proportional to the vapor pressure (P). To achieve higher cooling power at low temperatures, a $^3$He evaporation refrigerator is preferred over that $^4$He. Further, the evaporation refrigerators are operated in closed cycles and hence there is no $^3$He consumption. This brings down the operational cost of a $^3$He refrigerator similar to that of a $^4$He refrigerator. Currently, $^4$He evaporation refrigerators are operating at the Harvard University and the University of the Nevada Reno (UNR). The technical challenges of developing such refrigerators have already been overcome~\cite{Singh:PRL108:203201}.}. 

The molecules are first vaporized and cooled down to 3~K inside the first stage helium buffer gas cell~\footnote{The second stage of the cryogenic refrigerator (CRYOMECH:PT420) has a cooling power of 2~W at 4.2~K (http://www.cryomech.com/cryorefrigerators/pulse-tube/). As shown in the~\autoref{sec:appendix}, the total heat load on the second stage of the cryogenic refrigerator is much smaller than this cooling power.}. These precooled molecules are then extracted into the second stage where they are further cooled down to 0.5~K by collisions with buffer gas atoms. The heat load on the 0.5~K helium buffer gas cell is therefore reduced well below the cooling power of the $^3$He evaporation refrigerator at 0.5~K~\footnote{As shown in the~\autoref{sec:appendix}, the calculated heat load on the $^3$He evaporatiion refrigerator stays below 1~mW, when the beam source is operated with a repetition rate of 20~Hz}.

Our calculations demonstrate that for typically used buffer gas cell parameters and helium flow rates, the loss of the molecules inside the buffer gas cells can be minimized and hence a maximum number of the cold molecules can be extracted from the buffer gas cells into the molecular beam. The flux of the cold molecules produced from this source is calculated using the experimental values of the number of the CaF molecules produced by laser ablation and extraction efficiency the molecules from the buffer gas cell. On the other hand, the forward and transverse velocities of the molecules are calculated from the buffer gas temperature. The relevant equations are simplified and can be utilized straightforwardly for investigating the properties of buffer gas-cooled beams of any small molecule in the gas phase~\cite{Hutzler:CR112:4803,Kozyryev:PRL118:173201,Piskorski:CPC15:3800}.

\section{First stage buffer gas cooling of the molecules}
\label{sec:first}
\begin{figure}
   \centering%
   \includegraphics[width=\linewidth]{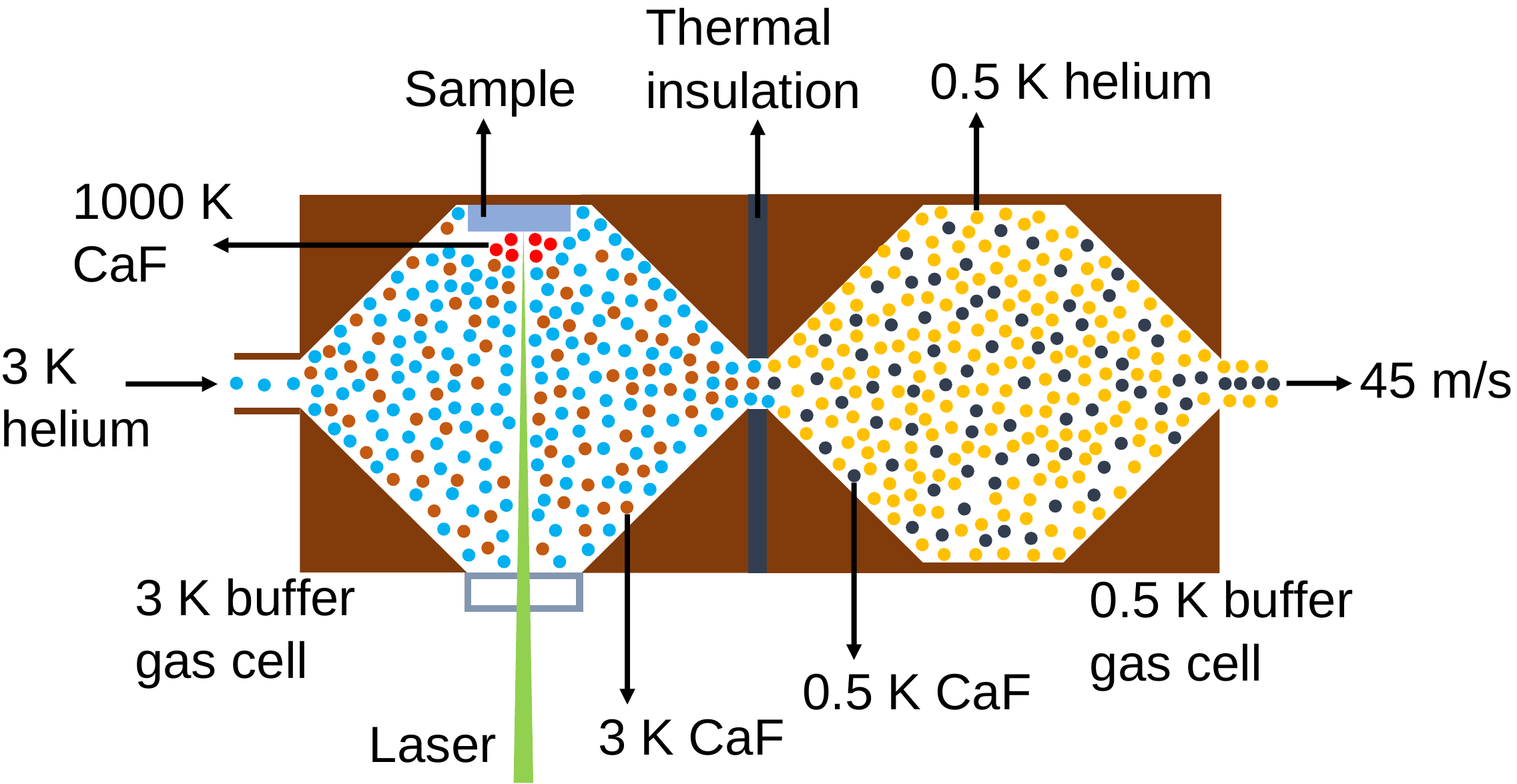}
   \caption{A schematic diagram of the two-stage helium buffer gas-cooled molecular beam source as discussed in the text. The 	 
   molecules are produced by laser ablation of a solid sample (CaF$_2$) and subsequently cooled down to 3~K in a 4~cm long and 5~cm 
   diameter first stage helium buffer gas cell. The molecules then extracted out through an exit aperture of 4~mm diameter and enter 
   into a 3~cm long and 5~cm diameter 0.5~K helium buffer gas cell through an entrance aperture of 5~mm diameter. The molecules then 
   cooled down to 0.5~K by collisions with helium atoms. The cold molecules then extracted out into the high vacuum through a 6~mm 
   diameter aperture and form a molecular beam. The buffer gas density inside the cell can be optimized by varying helium flow rate, 
   cell length, and the diameter of the cell exit aperture. The buffer gas pressure as low as 10$^{-6}$~mbar can be maintained inside 
   the vacuum chamber using a charcoal cryogenic sorption pump~\citep{Singh:PRA97:032704}.}
   \label{fig:cell}
\end{figure}

As shown in~\autoref{fig:cell}, the gas-phase molecules are produced inside the first stage helium buffer gas cell via laser ablation of a solid CaF$_2$ precursor and subsequently cooled down to 3~K by collisions with helium atoms. These precooled molecules are first extracted into the second stage helium buffer gas cell and subsequently cooled down to 0.5~K by collisions with cold helium atoms. To extract the maximum number of the molecules into the 0.5~K buffer gas cell, the 3~K buffer gas cell is operated in the hydrodynamic entrainment regime. As discussed in~\autoref{sec:Sim}, the buffer gas cells with conical geometries are used for circumventing the formation of the vortices in the helium flow. Thus, the diffusion losses due to the trapping of the molecules in the vortices are suppressed and hence the flux of the buffer gas-cooled molecules is further enhanced~\citep{Singh:PRA97:032704}.  

The initial temperature of the CaF is assumed to be 1000 K. Using the hard sphere collision model, the number of collisions required to thermalize the CaF molecules with 3~K $^4$He buffer gas can be estimated from~\autoref{eq1}~\cite{Hutzler:CR112:4803}.  
\begin{equation} \label{eq1}
T_\text{C} = T + (T_\text{i} - T)\exp{(\frac{-C}{\kappa})}
\end{equation}  
Where $T_\text{C}$ is the temperature of the molecules after $C$ collisions with buffer gas atoms, $T$ is the buffer gas temperature, $T_\text{i}$ is the initial temperature of the molecules, $\kappa = \frac{(M + m)^2}{2Mm}$ , and $M$ and $m$ are the masses of the molecule and buffer gas atoms respectively. For thermalizing the CaF molecules with 3~K $^4$He buffer gas atoms only 100 collisions are sufficient.

The length ($l$) required to thermalize the molecules with the buffer gas is related to the number of collisions ($C$) and the mean free path ($\lambda = 1/(\sqrt{2}\sigma n)$, by the~\autoref{eq2}.
\begin{equation} \label{eq2}
\begin{split}
l& = C\lambda\\
 & = C/(\sqrt{2}\sigma n)
\end{split}
\end{equation}
Where $\sigma$ is the collision cross-section of molecules with the buffer gas atoms and $n$ is the buffer gas density. 

In steady-state, the buffer gas density inside the cell is related to the buffer gas flow rate $f$ by the~\autoref{eq3}~\cite{Hutzler:CR112:4803}.
\begin{equation} \label{eq3}
n\approx\frac{4 f}{A\bar{v}}
\end{equation}
Where $A$ is the area of the cell exit aperture. The $\bar{v}$ is mean thermal velocity of the buffer gas atoms and is given by the following equation.
\begin{equation} \label{eq4}
\bar{v} = \sqrt{\frac{8 k_\text{B} T}{\pi m}}
\end{equation}
The $k_\text{B}$ is the Boltzmann constant. Using \autoref{eq3} and \autoref{eq4}, the $^4$He buffer gas density $n_\text{He}$ inside the first stage of the buffer gas cell can be estimated from~
\autoref{eq5}.
\begin{equation} \label{eq5}
\begin{split}
n_\text{He}&\approx 1.2\times10^{15}(\frac{f}{1~\text{SCCM}})(\frac{12~\text{mm}
^2}{A})\\
&\times (\frac{3~\text{K}}{T})^{1/2}(\frac{m}{6.6\times 10^{-27}~
\text{Kg}})^{1/2}~\text{cm}^{-3}
\end{split}
\end{equation}   
Where SCCM stands for standard cubic centimeter per minute. For the buffer gas flow rate of 10~SCCM and the cell with an exit aperture of 4~mm diameter, the calculated buffer gas density inside the cell is $\approx$ 1.2$\times$10$^{16}$~cm$^{-3}$. Using \autoref{eq2} and \autoref{eq5}, the thermalization length of the molecules with the $^4$He buffer gas can be given by the~\autoref{eq6}.
\begin{equation} \label{eq6}
\begin{split}
l&\approx 6\times(\frac{C}{100})(\frac{10^{-14}~\text{cm}^2}{\sigma})(\frac{1~
\text{SCCM}}{f})\\
&\times (\frac{A}{12~\text{mm}^2})(\frac{T}{3~\text{K}})^{1/2}
(\frac{6.6\times 10^{-27}~\text{Kg}}{m})^{1/2}~\text{cm}
\end{split}
\end{equation}
Here $\sigma_\text{CaF-He}$ is assumed to be 10$^{-14}$~cm$^2$~\cite{Hutzler:CR112:4803}. As expected the thermalization length decreases with increase in flow rate. For the helium flow rate of 12~SCCM, the length ($l$) required to thermalize the molecules with the 3~K helium buffer gas is 0.5~cm. Thus, a few cm long helium buffer gas cell should be sufficient for the first stage buffer gas cooling of molecules.
The number of the molecules extracted from the buffer gas cell into the molecular beam are characterized by the cell extraction parameter ($\gamma_\text{e} = \tau_\text{diff}/\tau_\text{pump}$), where $\tau_\text{diff}$ is the diffusion time of the molecules to the cold cell walls and $\tau_\text{pump}$ is the typical time spent by the molecules inside the buffer gas cell. For our first stage buffer gas cell, the $\gamma_\text{e}$ can be estimated from the~\autoref{eq7}~\cite{Hutzler:CR112:4803}. 
\begin{equation} \label{eq7}
\begin{split}
\gamma_\text{e}& = \frac{\sigma f}{L \bar{v}}\approx 1.1\times(\frac{\sigma}
{10^{-14}~\text{cm}^2})(\frac{f}{12~\text{SCCM}})\\
&\times (\frac{4~\text{cm}}{L})(\frac{3~\text{K}}{T})^{1/2}(\frac{m}
{6.6\times 10^{-27}~\text{Kg}})^{1/2}
\end{split}
\end{equation}
For our helium buffer gas cell length ($L$) of 4~cm, and buffer gas flow rate of 12~SCCM, $\gamma_\text{e}\approx$ 1.1. Thus, the cell will operate in hydrodynamic entrainment regime ($\gamma_\text{e} \geq$~1). 

It has been observed in previous experiments that the number of CaF molecules as high as $5\times$10$^{13}$ can be produced in each ablation pulse~\cite{Maussang:PRL94:123002}. In addition to this, the extraction of over 40$\%$ of the buffer gas-cooled molecules from the buffer gas cell into the molecular beam has been observed~\cite{Patterson:JCP126:154307}, when the buffer gas cell is operated in hydrodynamic entrainment regime. With this extraction efficiency, the number of the buffer gas-cooled molecules transmitted through the 3~K buffer gas cell exit aperture are $\approx$ 2$\times$ 10$^{13}$ per pulse. This extraction of the molecules can be further increased by using an optimized buffer gas cell geometry~\cite{Singh:PRA97:032704}.

Owing to their larger mass, the mean thermal velocity of the CaF molecules inside the buffer gas cell is smaller than that of the buffer gas atoms. At the cell exit, the collisions of the buffer gas atoms boost the mean forward velocity of the molecules. The number of collision between the molecules and buffer gas atoms can be estimated from~\autoref{eq8}~\cite{Hutzler:CR112:4803}.
\begin{equation} \label{eq8}
\frac{d}{\lambda}~\approx~\frac{R_\text{e}}{2}
\end{equation} 
Where $d$ is the diameter of cell exit aperture and $R_\text{e}$ is the Reynolds number. Using~\autoref{eq2},~\autoref{eq3}, and~\autoref{eq8}, the Reynolds number can be given by the following equation.
\begin{equation} \label{eq9}
\begin{split}
R_\text{e}&\approx\frac{10\sigma f d}{A \bar{v}}\approx 
8.6\times(\frac{\sigma}{10^{-14}~\text{cm}^2})(\frac{f}{1~\text{SCCM}})\\
&\times (\frac{4~\text{mm}}{d})(\frac{3~\text{K}}{T})^{1/2}(\frac{m}
{6.6\times 10^{-27}~\text{Kg}})^{1/2}
\end{split}
\end{equation}
The Reynolds number increases with buffer gas flow rate. For the helium flow rate of 12~SCCM, the Reynolds number $\approx$ 100. At the cell exit, the number of collisions of molecules with buffer gas are $\approx$ 50. In each collision with buffer gas atom, the change in the forward momentum of the molecules $\approx m\bar{v}$. Thus, the corresponding change in the forward velocity ($R_\text{e}m\bar{v}/{2M}$) of the molecules can be estimated from~\autoref{eq10}.
\begin{equation} \label{eq10}
\begin{split}
\Delta v_{\text{mol},\parallel}&\approx 40\times(\frac{\sigma}{10^{-14}~
\text{cm}^2})(\frac{f}{1~\text{SCCM}})(\frac{4~\text{mm}}{d})\\
&\times(\frac{m}{4~\text{amu}})(\frac{59~\text{amu}}{M})~\text{m/s}
\end{split}
\end{equation}
The number of collisions between the molecules and the buffer gas atoms increase with buffer gas flow rate. At 12~SCCM helium flow rate, the forward velocity of the molecules is expected to be boosted to the forward velocity ($v_\text{f} = \sqrt{\frac{2k_\text{B}T}{m}}$) of the buffer gas atoms, i.e. 110~m/s. 

Further, the transverse velocity of the buffer gas inside the cell ($v_\text{cell}$) can be approximated by~\autoref{eq11}~\cite{Hutzler:PCCP13:18976}.
\begin{equation} \label{eq11}
\begin{split}
v_\text{cell}&\approx \bar{v}\times(\frac{d}{d_\text{cell}})^2\approx 
0.8\times(\frac{d}{4~\text{mm}})^2(\frac{50~\text{mm}}{d_\text{cell}})^2\\
&\times (\frac{T}{3~\text{K}})^{1/2}(\frac{6.6\times 10^{-27}~\text{Kg}}
{m})^{1/2}~\text{m/s} 
\end{split}
\end{equation}
Where $d_\text{cell}$ is the cell diameter. For the helium buffer gas cell with 50~mm diameter, $v_\text{cell} \approx$ 0.8~m/s. The change in the transverse velocity ($R_\text{e} mv_\text{cell}/2M$) of the molecules due to the collisions with buffer gas atoms at the cell exit can be estimated from~\autoref{eq12}. 
\begin{equation} \label{eq12}
\begin{split}
\Delta v_{\text{mol},\perp}&\approx 0.2\times(\frac{\sigma}{10^{-14}~\text{cm}
^2})(\frac{f}{1~\text{SCCM}})(\frac{d}{4~\text{mm}})\\
&\times(\frac{50~\text{mm}}{d_\text{cell}})^2(\frac{m}{4~\text{amu}})
(\frac{59~\text{amu}}{M})~\text{m/s}
\end{split}
\end{equation}
The change in the transverse velocity of the molecules also increases with increase in the buffer gas flow rate. At 12~SCCM, the change in transverse velocity $\approx$ 2.4~m/s. Assuming a 4~mm initial beam spread (size of the exit aperture), the calculated transverse molecular beam spread 1~cm away from the first stage buffer gas cell exit $\approx$ 4.4~mm~
\footnote{The time taken by the beam in traveling 1~cm is 0.09~ms. The transverse distance traveled by the beam during this time is $\pm$ 0.2 mm.}. Thus, using a 5~mm diameter entrance aperture, all the precooled CaF molecules transmitted from the first stage buffer gas cell can be extracted into the second stage buffer gas cell. The distance between the cells can indeed be optimized for extracting the maximum number of the molecules from first stage buffer gas cell into the second stage buffer gas cell. 

\section{Second stage buffer gas cooling of the molecules}
\label{sec:second}
To suppress the diffusion losses and hence extracting the maximum number of the molecules from the cell into the molecular beam, the 0.5~K helium buffer gas cell is also operated in the hydrodynamic entrainment regime. From~\autoref{eq1}, the number of collisions of the CaF molecules with helium buffer gas atoms required to cool them from 3~K down to 0.5~K is 50. For a helium flow rate of 5~SCCM and the cell exit aperture of 6~mm diameter, the thermalization length of the molecules with the 0.5~K helium buffer gas calculated using~\autoref{eq6} is 0.6~cm. For this flow, the helium buffer gas density estimated from~\autoref{eq5} is $\approx$ 6$\times$ 10$^{15}$~cm$^{-3}$.

With a helium buffer gas length of 3~cm, the calculated extraction parameter is $\gamma_e\approx$ 1.5. Thus, the cell will operate in hydrodynamic entrainment regime and over 40$\%$ of the buffer gas-cooled molecules out of the 2$\times$ 10$^{13}$ can be extracted from the cell into the beam~\cite{Patterson:JCP126:154307}. Thus, the flux of produced CaF molecules per pulse is $\approx$ 8$\times$10$^{12}$. It is worth mentioning that the flux of the cold atoms and molecules produced from a buffer gas-cooled beam source largely depends on the vaporization yield. It has been observed that the high purity (99.9\%) solid precursors yield higher fluxes of cold atoms and molecules as compared to the smaller purity (95\%) precursors~\cite{Maussang:PRL94:123002,Singh:PRL108:203201}. 

On the other hand, the mean forward velocity of the molecules in the beam depends upon the buffer gas temperature and the number of the collisions of molecules with the buffer gas atoms at the cell exit aperture. For the second stage helium buffer gas cell parameters, the estimated Reynolds number from~\autoref{eq9} is 70. The number of collisions of molecules with buffer gas atoms is $\approx$ 35. From~\autoref{eq10}, it is clear that these collisions are sufficient to boost the forward velocity of the molecular beam to the forward velocity ($v_\text{f} = \sqrt{\frac{2k_\text{B}T}{m}}$) of the buffer gas atoms. At 0.5~K, the CaF will acquire a mean forward velocity of 45~m/s.
 
As discussed in~\autoref{sec:PDF}, at 0.5~K, the Maxwell-Boltzmann PDF of the helium atoms moving with speeds $\leq$~5~m/s is 0.001. In the hydrodynamic entrainment regime, the velocity distribution of the molecules is similar to that of the buffer gas atoms. Thus, the flux of the CaF moving with speeds $\leq$~5~m/s is 8$\times$10$^{9}$ molecules per pulse. These very slow and intense beams of the cold molecules can be directly loaded into a magneto-optical trap and hence paving the way for efficient laser cooling of the molecules~\cite{Anderegg:PRL119:103201,Williams:PRL120:163201}. In addition to this, these very slow and high flux beams of the molecules can be utilized for investigating the sympathetic cooling of the buffer gas-cooled molecules with ultracold atoms~\cite{Lim:PRA92:053419}. Furthermore, such very slow and intense beams of the buffer gas-cooled molecules can also be utilized for searching the electric dipole moment of the electron~\citep{ACME:Science343:269,Hudson:Nature473:493} and measurements of the parity-violating energy difference between enantiomers of the chiral molecules~\citep{QUACK:CPL132:147,Tokunaga:NJP19:053006}. 

\section{Conclusion}
\label{sec:conclusion}
In conclusion, a novel two-stage buffer gas-cooled beam source is introduced. The vaporization and precooling of the molecules inside the first stage 3~K helium buffer gas cell suppresses the heat load on the second stage and hence facilitates the operation of the second stage buffer gas cell at 0.5~K with the repetition rates over 20~Hz. At this temperature, the mean forward velocity and the flux of the CaF beam are calculated to be 45~m/s and 8$\times 10^{12}$ molecules per pulse respectively, when both buffer gas cells are operated in the hydrodynamic entrainment regime. Using this flux and Maxwell-Boltzmann PDF at 0.5~K, the number of the molecules moving with speeds $\leq$~5~m/s is calculated to be 8$\times 10^{9}$. These very slow and intense beams of the cold molecules produced from our two-stage buffer gas-cooled beam source can be utilized for a wide diversity of molecular physics experiments such as efficient magneto-optical trapping of the molecules and performing ultrahigh precision molecular spectroscopy.

\section{Acknowledgements}
\label{sec:acknowledgements}

This work is supported by the National Science Centre, Poland, Project No. 2017/25/B/ST2/00429. 

\bibliography{Radha}

\begin{thebibliography}{32}%
\makeatletter
\providecommand \@ifxundefined [1]{%
 \@ifx{#1\undefined}
}%
\providecommand \@ifnum [1]{%
 \ifnum #1\expandafter \@firstoftwo
 \else \expandafter \@secondoftwo
 \fi
}%
\providecommand \@ifx [1]{%
 \ifx #1\expandafter \@firstoftwo
 \else \expandafter \@secondoftwo
 \fi
}%
\providecommand \natexlab [1]{#1}%
\providecommand \enquote  [1]{``#1''}%
\providecommand \bibnamefont  [1]{#1}%
\providecommand \bibfnamefont [1]{#1}%
\providecommand \citenamefont [1]{#1}%
\providecommand \href@noop [0]{\@secondoftwo}%
\providecommand \href [0]{\begingroup \@sanitize@url \@href}%
\providecommand \@href[1]{\@@startlink{#1}\@@href}%
\providecommand \@@href[1]{\endgroup#1\@@endlink}%
\providecommand \@sanitize@url [0]{\catcode `\\12\catcode `\$12\catcode
  `\&12\catcode `\#12\catcode `\^12\catcode `\_12\catcode `\%12\relax}%
\providecommand \@@startlink[1]{}%
\providecommand \@@endlink[0]{}%
\providecommand \url  [0]{\begingroup\@sanitize@url \@url }%
\providecommand \@url [1]{\endgroup\@href {#1}{\urlprefix }}%
\providecommand \urlprefix  [0]{URL }%
\providecommand \Eprint [0]{\href }%
\providecommand \doibase [0]{http://dx.doi.org/}%
\providecommand \selectlanguage [0]{\@gobble}%
\providecommand \bibinfo  [0]{\@secondoftwo}%
\providecommand \bibfield  [0]{\@secondoftwo}%
\providecommand \translation [1]{[#1]}%
\providecommand \BibitemOpen [0]{}%
\providecommand \bibitemStop [0]{}%
\providecommand \bibitemNoStop [0]{.\EOS\space}%
\providecommand \EOS [0]{\spacefactor3000\relax}%
\providecommand \BibitemShut  [1]{\csname bibitem#1\endcsname}%
\let\auto@bib@innerbib\@empty
\bibitem [{\citenamefont {Weinstein}\ \emph {et~al.}(1998)\citenamefont
  {Weinstein}, \citenamefont {deCarvalho}, \citenamefont {Guillet},
  \citenamefont {Friedrich},\ and\ \citenamefont
  {Doyle}}]{Weinstein:Nature395:148}%
  \BibitemOpen
  \bibfield  {author} {\bibinfo {author} {\bibfnamefont {Jonathan~D.}\
  \bibnamefont {Weinstein}}, \bibinfo {author} {\bibfnamefont {Robert}\
  \bibnamefont {deCarvalho}}, \bibinfo {author} {\bibfnamefont {Thierry}\
  \bibnamefont {Guillet}}, \bibinfo {author} {\bibfnamefont {Bretislav}\
  \bibnamefont {Friedrich}}, \ and\ \bibinfo {author} {\bibfnamefont {John~M.}\
  \bibnamefont {Doyle}},\ }\bibfield  {title} {\enquote {\bibinfo {title}
  {Magnetic trapping of calcium monohydride molecules at millikelvin
  temperatures},}\ }\href@noop {} {\bibfield  {journal} {\bibinfo  {journal}
  {Nature}\ }\textbf {\bibinfo {volume} {395}},\ \bibinfo {pages} {148--150}
  (\bibinfo {year} {1998})}\BibitemShut {NoStop}%
\bibitem [{\citenamefont {Hutzler}\ \emph {et~al.}(2012)\citenamefont
  {Hutzler}, \citenamefont {Lu},\ and\ \citenamefont
  {Doyle}}]{Hutzler:CR112:4803}%
  \BibitemOpen
  \bibfield  {author} {\bibinfo {author} {\bibfnamefont {N.~R.}\ \bibnamefont
  {Hutzler}}, \bibinfo {author} {\bibfnamefont {H.-I}\ \bibnamefont {Lu}}, \
  and\ \bibinfo {author} {\bibfnamefont {J.~M.}\ \bibnamefont {Doyle}},\
  }\bibfield  {title} {\enquote {\bibinfo {title} {The buffer gas beam: An
  intense, cold, and slow source for atoms and molecules},}\ }\href@noop {}
  {\bibfield  {journal} {\bibinfo  {journal} {Chem. Rev.}\ }\textbf {\bibinfo
  {volume} {112}},\ \bibinfo {pages} {4803} (\bibinfo {year}
  {2012})}\BibitemShut {NoStop}%
\bibitem [{\citenamefont {Singh}\ \emph {et~al.}(2012)\citenamefont {Singh},
  \citenamefont {Hardman}, \citenamefont {Tariq}, \citenamefont {Lu},
  \citenamefont {Ellis}, \citenamefont {Morrison},\ and\ \citenamefont
  {Weinstein}}]{Singh:PRL108:203201}%
  \BibitemOpen
  \bibfield  {author} {\bibinfo {author} {\bibfnamefont {Vijay}\ \bibnamefont
  {Singh}}, \bibinfo {author} {\bibfnamefont {Kyle~S.}\ \bibnamefont
  {Hardman}}, \bibinfo {author} {\bibfnamefont {Naima}\ \bibnamefont {Tariq}},
  \bibinfo {author} {\bibfnamefont {Mei-Ju}\ \bibnamefont {Lu}}, \bibinfo
  {author} {\bibfnamefont {Aja}\ \bibnamefont {Ellis}}, \bibinfo {author}
  {\bibfnamefont {Muir~J.}\ \bibnamefont {Morrison}}, \ and\ \bibinfo {author}
  {\bibfnamefont {Jonathan~D.}\ \bibnamefont {Weinstein}},\ }\bibfield  {title}
  {\enquote {\bibinfo {title} {Chemical reactions of atomic lithium and
  molecular calcium monohydride at 1 {K}},}\ }\href {\doibase
  10.1103/PhysRevLett.108.203201} {\bibfield  {journal} {\bibinfo  {journal}
  {Phys. Rev. Lett.}\ }\textbf {\bibinfo {volume} {108}},\ \bibinfo {pages}
  {203201} (\bibinfo {year} {2012})}\BibitemShut {NoStop}%
\bibitem [{\citenamefont {Anderegg}\ \emph {et~al.}(2017)\citenamefont
  {Anderegg}, \citenamefont {Augenbraun}, \citenamefont {Chae}, \citenamefont
  {Hemmerling}, \citenamefont {Hutzler}, \citenamefont {Ravi}, \citenamefont
  {Collopy}, \citenamefont {Ye}, \citenamefont {Ketterle},\ and\ \citenamefont
  {Doyle}}]{Anderegg:PRL119:103201}%
  \BibitemOpen
  \bibfield  {author} {\bibinfo {author} {\bibfnamefont {Lo\"{\i}c}\
  \bibnamefont {Anderegg}}, \bibinfo {author} {\bibfnamefont {Benjamin~L.}\
  \bibnamefont {Augenbraun}}, \bibinfo {author} {\bibfnamefont {Eunmi}\
  \bibnamefont {Chae}}, \bibinfo {author} {\bibfnamefont {Boerge}\ \bibnamefont
  {Hemmerling}}, \bibinfo {author} {\bibfnamefont {Nicholas~R.}\ \bibnamefont
  {Hutzler}}, \bibinfo {author} {\bibfnamefont {Aakash}\ \bibnamefont {Ravi}},
  \bibinfo {author} {\bibfnamefont {Alejandra}\ \bibnamefont {Collopy}},
  \bibinfo {author} {\bibfnamefont {Jun}\ \bibnamefont {Ye}}, \bibinfo {author}
  {\bibfnamefont {Wolfgang}\ \bibnamefont {Ketterle}}, \ and\ \bibinfo {author}
  {\bibfnamefont {John~M.}\ \bibnamefont {Doyle}},\ }\bibfield  {title}
  {\enquote {\bibinfo {title} {Radio frequency magneto-optical trapping of
  {CaF} with high density},}\ }\href {\doibase 10.1103/PhysRevLett.119.103201}
  {\bibfield  {journal} {\bibinfo  {journal} {Phys. Rev. Lett.}\ }\textbf
  {\bibinfo {volume} {119}},\ \bibinfo {pages} {103201} (\bibinfo {year}
  {2017})}\BibitemShut {NoStop}%
\bibitem [{\citenamefont {Shuman}\ \emph {et~al.}(2010)\citenamefont {Shuman},
  \citenamefont {Barry},\ and\ \citenamefont {DeMille}}]{Shuman:Nature467:820}%
  \BibitemOpen
  \bibfield  {author} {\bibinfo {author} {\bibfnamefont {E.~S.}\ \bibnamefont
  {Shuman}}, \bibinfo {author} {\bibfnamefont {J.~F.}\ \bibnamefont {Barry}}, \
  and\ \bibinfo {author} {\bibfnamefont {D.}~\bibnamefont {DeMille}},\
  }\bibfield  {title} {\enquote {\bibinfo {title} {Laser cooling of a diatomic
  molecule},}\ }\href@noop {} {\bibfield  {journal} {\bibinfo  {journal}
  {Nature}\ }\textbf {\bibinfo {volume} {467}},\ \bibinfo {pages} {820–823}
  (\bibinfo {year} {2010})}\BibitemShut {NoStop}%
\bibitem [{\citenamefont {Kozyryev}\ \emph {et~al.}(2017)\citenamefont
  {Kozyryev}, \citenamefont {Baum}, \citenamefont {Matsuda}, \citenamefont
  {Augenbraun}, \citenamefont {Anderegg}, \citenamefont {Sedlack},\ and\
  \citenamefont {Doyle}}]{Kozyryev:PRL118:173201}%
  \BibitemOpen
  \bibfield  {author} {\bibinfo {author} {\bibfnamefont {Ivan}\ \bibnamefont
  {Kozyryev}}, \bibinfo {author} {\bibfnamefont {Louis}\ \bibnamefont {Baum}},
  \bibinfo {author} {\bibfnamefont {Kyle}\ \bibnamefont {Matsuda}}, \bibinfo
  {author} {\bibfnamefont {Benjamin~L.}\ \bibnamefont {Augenbraun}}, \bibinfo
  {author} {\bibfnamefont {Loic}\ \bibnamefont {Anderegg}}, \bibinfo {author}
  {\bibfnamefont {Alexander~P.}\ \bibnamefont {Sedlack}}, \ and\ \bibinfo
  {author} {\bibfnamefont {John~M.}\ \bibnamefont {Doyle}},\ }\bibfield
  {title} {\enquote {\bibinfo {title} {Sisyphus laser cooling of a polyatomic
  molecule},}\ }\href {\doibase 10.1103/PhysRevLett.118.173201} {\bibfield
  {journal} {\bibinfo  {journal} {Phys. Rev. Lett.}\ }\textbf {\bibinfo
  {volume} {118}},\ \bibinfo {pages} {173201} (\bibinfo {year}
  {2017})}\BibitemShut {NoStop}%
\bibitem [{\citenamefont {Williams}\ \emph {et~al.}(2018)\citenamefont
  {Williams}, \citenamefont {Caldwell}, \citenamefont {Fitch}, \citenamefont
  {Truppe}, \citenamefont {Rodewald}, \citenamefont {Hinds}, \citenamefont
  {Sauer},\ and\ \citenamefont {Tarbutt}}]{Williams:PRL120:163201}%
  \BibitemOpen
  \bibfield  {author} {\bibinfo {author} {\bibfnamefont {H.~J.}\ \bibnamefont
  {Williams}}, \bibinfo {author} {\bibfnamefont {L.}~\bibnamefont {Caldwell}},
  \bibinfo {author} {\bibfnamefont {N.~J.}\ \bibnamefont {Fitch}}, \bibinfo
  {author} {\bibfnamefont {S.}~\bibnamefont {Truppe}}, \bibinfo {author}
  {\bibfnamefont {J.}~\bibnamefont {Rodewald}}, \bibinfo {author}
  {\bibfnamefont {E.~A.}\ \bibnamefont {Hinds}}, \bibinfo {author}
  {\bibfnamefont {B.~E.}\ \bibnamefont {Sauer}}, \ and\ \bibinfo {author}
  {\bibfnamefont {M.~R.}\ \bibnamefont {Tarbutt}},\ }\bibfield  {title}
  {\enquote {\bibinfo {title} {Magnetic trapping and coherent control of
  laser-cooled molecules},}\ }\href {\doibase 10.1103/PhysRevLett.120.163201}
  {\bibfield  {journal} {\bibinfo  {journal} {Phys. Rev. Lett.}\ }\textbf
  {\bibinfo {volume} {120}},\ \bibinfo {pages} {163201} (\bibinfo {year}
  {2018})}\BibitemShut {NoStop}%
\bibitem [{\citenamefont {Zhelyazkova}\ \emph {et~al.}(2014)\citenamefont
  {Zhelyazkova}, \citenamefont {Cournol}, \citenamefont {Wall}, \citenamefont
  {Matsushima}, \citenamefont {Hudson}, \citenamefont {Hinds}, \citenamefont
  {Tarbutt},\ and\ \citenamefont {Sauer}}]{Zhelyazkova:PRA89:053416}%
  \BibitemOpen
  \bibfield  {author} {\bibinfo {author} {\bibfnamefont {V.}~\bibnamefont
  {Zhelyazkova}}, \bibinfo {author} {\bibfnamefont {A.}~\bibnamefont
  {Cournol}}, \bibinfo {author} {\bibfnamefont {T.~E.}\ \bibnamefont {Wall}},
  \bibinfo {author} {\bibfnamefont {A.}~\bibnamefont {Matsushima}}, \bibinfo
  {author} {\bibfnamefont {J.~J.}\ \bibnamefont {Hudson}}, \bibinfo {author}
  {\bibfnamefont {E.~A.}\ \bibnamefont {Hinds}}, \bibinfo {author}
  {\bibfnamefont {M.~R.}\ \bibnamefont {Tarbutt}}, \ and\ \bibinfo {author}
  {\bibfnamefont {B.~E.}\ \bibnamefont {Sauer}},\ }\bibfield  {title} {\enquote
  {\bibinfo {title} {Laser cooling and slowing of {CaF} molecules},}\ }\href
  {\doibase 10.1103/PhysRevA.89.053416} {\bibfield  {journal} {\bibinfo
  {journal} {Phys. Rev. A}\ }\textbf {\bibinfo {volume} {89}},\ \bibinfo
  {pages} {053416} (\bibinfo {year} {2014})}\BibitemShut {NoStop}%
\bibitem [{\citenamefont {Lavert-Ofir}\ \emph {et~al.}(2011)\citenamefont
  {Lavert-Ofir}, \citenamefont {Gersten}, \citenamefont {Henson}, \citenamefont
  {Shani}, \citenamefont {David}, \citenamefont {Narevicius},\ and\
  \citenamefont {Narevicius}}]{Ofir:NJP13:103030}%
  \BibitemOpen
  \bibfield  {author} {\bibinfo {author} {\bibfnamefont {Etay}\ \bibnamefont
  {Lavert-Ofir}}, \bibinfo {author} {\bibfnamefont {Sasha}\ \bibnamefont
  {Gersten}}, \bibinfo {author} {\bibfnamefont {Alon~B}\ \bibnamefont
  {Henson}}, \bibinfo {author} {\bibfnamefont {Itamar}\ \bibnamefont {Shani}},
  \bibinfo {author} {\bibfnamefont {Liron}\ \bibnamefont {David}}, \bibinfo
  {author} {\bibfnamefont {Julia}\ \bibnamefont {Narevicius}}, \ and\ \bibinfo
  {author} {\bibfnamefont {Edvardas}\ \bibnamefont {Narevicius}},\ }\bibfield
  {title} {\enquote {\bibinfo {title} {A moving magnetic trap decelerator: a
  new source of cold atoms and molecules},}\ }\href
  {http://stacks.iop.org/1367-2630/13/i=10/a=103030} {\bibfield  {journal}
  {\bibinfo  {journal} {New J. Phys.}\ }\textbf {\bibinfo {volume} {13}},\
  \bibinfo {pages} {103030} (\bibinfo {year} {2011})}\BibitemShut {NoStop}%
\bibitem [{\citenamefont {McArd}\ \emph {et~al.}()\citenamefont {McArd},
  \citenamefont {Mizouri}, \citenamefont {Walker}, \citenamefont {Singh},
  \citenamefont {Krohn}, \citenamefont {Hinds},\ and\ \citenamefont
  {Carty}}]{McArd:MTZD:inprep}%
  \BibitemOpen
  \bibfield  {author} {\bibinfo {author} {\bibfnamefont {Lewis~A.}\
  \bibnamefont {McArd}}, \bibinfo {author} {\bibfnamefont {Arin}\ \bibnamefont
  {Mizouri}}, \bibinfo {author} {\bibfnamefont {Paul~A.}\ \bibnamefont
  {Walker}}, \bibinfo {author} {\bibfnamefont {Vijay}\ \bibnamefont {Singh}},
  \bibinfo {author} {\bibfnamefont {Ulrich}\ \bibnamefont {Krohn}}, \bibinfo
  {author} {\bibfnamefont {Ed~A.}\ \bibnamefont {Hinds}}, \ and\ \bibinfo
  {author} {\bibfnamefont {David}\ \bibnamefont {Carty}},\ }\bibfield  {title}
  {\enquote {\bibinfo {title} {A moving-trap zeeman decelerator},}\ }\href
  {https://arxiv.org/abs/1807.10648} {\bibfield  {journal} {\bibinfo  {journal}
  {preprint at}\ }}\Eprint {http://arxiv.org/abs/1807.10648} {arXiv:1807.10648}
  \BibitemShut {NoStop}%
\bibitem [{\citenamefont {Quintero-P\'erez}\ \emph {et~al.}(2013)\citenamefont
  {Quintero-P\'erez}, \citenamefont {Jansen}, \citenamefont {Wall},
  \citenamefont {van~den Berg}, \citenamefont {Hoekstra},\ and\ \citenamefont
  {Bethlem}}]{Perez:PRL110:133003}%
  \BibitemOpen
  \bibfield  {author} {\bibinfo {author} {\bibfnamefont {Marina}\ \bibnamefont
  {Quintero-P\'erez}}, \bibinfo {author} {\bibfnamefont {Paul}\ \bibnamefont
  {Jansen}}, \bibinfo {author} {\bibfnamefont {Thomas~E.}\ \bibnamefont
  {Wall}}, \bibinfo {author} {\bibfnamefont {Joost~E.}\ \bibnamefont {van~den
  Berg}}, \bibinfo {author} {\bibfnamefont {Steven}\ \bibnamefont {Hoekstra}},
  \ and\ \bibinfo {author} {\bibfnamefont {Hendrick~L.}\ \bibnamefont
  {Bethlem}},\ }\bibfield  {title} {\enquote {\bibinfo {title} {Static trapping
  of polar molecules in a traveling wave decelerator},}\ }\href {\doibase
  10.1103/PhysRevLett.110.133003} {\bibfield  {journal} {\bibinfo  {journal}
  {Phys. Rev. Lett.}\ }\textbf {\bibinfo {volume} {110}},\ \bibinfo {pages}
  {133003} (\bibinfo {year} {2013})}\BibitemShut {NoStop}%
\bibitem [{\citenamefont {Hudson}\ \emph {et~al.}(2011)\citenamefont {Hudson},
  \citenamefont {Kara}, \citenamefont {Smallman}, \citenamefont {Sauer},
  \citenamefont {Tarbutt},\ and\ \citenamefont {Hinds}}]{Hudson:Nature473:493}%
  \BibitemOpen
  \bibfield  {author} {\bibinfo {author} {\bibfnamefont {J.~J.}\ \bibnamefont
  {Hudson}}, \bibinfo {author} {\bibfnamefont {D.~M.}\ \bibnamefont {Kara}},
  \bibinfo {author} {\bibfnamefont {I.~J.}\ \bibnamefont {Smallman}}, \bibinfo
  {author} {\bibfnamefont {B.~E.}\ \bibnamefont {Sauer}}, \bibinfo {author}
  {\bibfnamefont {M.~R.}\ \bibnamefont {Tarbutt}}, \ and\ \bibinfo {author}
  {\bibfnamefont {E.~A.}\ \bibnamefont {Hinds}},\ }\bibfield  {title} {\enquote
  {\bibinfo {title} {Improved measurement of the shape of the electron},}\
  }\href@noop {} {\bibfield  {journal} {\bibinfo  {journal} {Nature}\ }\textbf
  {\bibinfo {volume} {473}},\ \bibinfo {pages} {493–496} (\bibinfo {year}
  {2011})}\BibitemShut {NoStop}%
\bibitem [{\citenamefont {Collaboration}\ \emph {et~al.}(2014)\citenamefont
  {Collaboration}, \citenamefont {Baron}, \citenamefont {Campbell},
  \citenamefont {DeMille}, \citenamefont {Doyle}, \citenamefont {Gabrielse},
  \citenamefont {Gurevich}, \citenamefont {Hess}, \citenamefont {Hutzler},
  \citenamefont {Kirilov}, \citenamefont {Kozyryev}, \citenamefont {O'Leary},
  \citenamefont {Panda}, \citenamefont {Petrik}, \citenamefont {Spaun},
  \citenamefont {Vutha},\ and\ \citenamefont {West}}]{ACME:Science343:269}%
  \BibitemOpen
  \bibfield  {author} {\bibinfo {author} {\bibfnamefont {ACME}\ \bibnamefont
  {Collaboration}}, \bibinfo {author} {\bibfnamefont {J.}~\bibnamefont
  {Baron}}, \bibinfo {author} {\bibfnamefont {W.~C.}\ \bibnamefont {Campbell}},
  \bibinfo {author} {\bibfnamefont {D.}~\bibnamefont {DeMille}}, \bibinfo
  {author} {\bibfnamefont {J.~M.}\ \bibnamefont {Doyle}}, \bibinfo {author}
  {\bibfnamefont {G.}~\bibnamefont {Gabrielse}}, \bibinfo {author}
  {\bibfnamefont {Y.~V.}\ \bibnamefont {Gurevich}}, \bibinfo {author}
  {\bibfnamefont {P.~W.}\ \bibnamefont {Hess}}, \bibinfo {author}
  {\bibfnamefont {N.~R.}\ \bibnamefont {Hutzler}}, \bibinfo {author}
  {\bibfnamefont {E.}~\bibnamefont {Kirilov}}, \bibinfo {author} {\bibfnamefont
  {I.}~\bibnamefont {Kozyryev}}, \bibinfo {author} {\bibfnamefont {B.~R.}\
  \bibnamefont {O'Leary}}, \bibinfo {author} {\bibfnamefont {C.~D.}\
  \bibnamefont {Panda}}, \bibinfo {author} {\bibfnamefont {E.~S.}\ \bibnamefont
  {Petrik}}, \bibinfo {author} {\bibfnamefont {B.}~\bibnamefont {Spaun}},
  \bibinfo {author} {\bibfnamefont {A.~C.}\ \bibnamefont {Vutha}}, \ and\
  \bibinfo {author} {\bibfnamefont {A.~D.}\ \bibnamefont {West}},\ }\bibfield
  {title} {\enquote {\bibinfo {title} {Order of magnitude smaller limit on the
  electric dipole moment of the electron},}\ }\href@noop {} {\bibfield
  {journal} {\bibinfo  {journal} {Science}\ }\textbf {\bibinfo {volume}
  {343}},\ \bibinfo {pages} {269--272} (\bibinfo {year} {2014})}\BibitemShut
  {NoStop}%
\bibitem [{\citenamefont {Quack}(1986)}]{QUACK:CPL132:147}%
  \BibitemOpen
  \bibfield  {author} {\bibinfo {author} {\bibfnamefont {Martin}\ \bibnamefont
  {Quack}},\ }\bibfield  {title} {\enquote {\bibinfo {title} {On the
  measurement of the parity violating energy difference between enantiomers},}\
  }\href {\doibase https://doi.org/10.1016/0009-2614(86)80098-7} {\bibfield
  {journal} {\bibinfo  {journal} {Chem. Phys. Lett.}\ }\textbf {\bibinfo
  {volume} {132}},\ \bibinfo {pages} {147 -- 153} (\bibinfo {year}
  {1986})}\BibitemShut {NoStop}%
\bibitem [{\citenamefont {Tokunaga}\ \emph {et~al.}(2017)\citenamefont
  {Tokunaga}, \citenamefont {Hendricks}, \citenamefont {Tarbutt},\ and\
  \citenamefont {Darquié}}]{Tokunaga:NJP19:053006}%
  \BibitemOpen
  \bibfield  {author} {\bibinfo {author} {\bibfnamefont {S~K}\ \bibnamefont
  {Tokunaga}}, \bibinfo {author} {\bibfnamefont {R~J}\ \bibnamefont
  {Hendricks}}, \bibinfo {author} {\bibfnamefont {M~R}\ \bibnamefont
  {Tarbutt}}, \ and\ \bibinfo {author} {\bibfnamefont {B}~\bibnamefont
  {Darquié}},\ }\bibfield  {title} {\enquote {\bibinfo {title}
  {High-resolution mid-infrared spectroscopy of buffer-gas-cooled
  methyltrioxorhenium molecules},}\ }\href
  {http://stacks.iop.org/1367-2630/19/i=5/a=053006} {\bibfield  {journal}
  {\bibinfo  {journal} {New J. Phys.}\ }\textbf {\bibinfo {volume} {19}},\
  \bibinfo {pages} {053006} (\bibinfo {year} {2017})}\BibitemShut {NoStop}%
\bibitem [{\citenamefont {Kozyryev}\ and\ \citenamefont
  {Hutzler}(2017)}]{Kozyryev:PRL119:133002}%
  \BibitemOpen
  \bibfield  {author} {\bibinfo {author} {\bibfnamefont {Ivan}\ \bibnamefont
  {Kozyryev}}\ and\ \bibinfo {author} {\bibfnamefont {Nicholas~R.}\
  \bibnamefont {Hutzler}},\ }\bibfield  {title} {\enquote {\bibinfo {title}
  {Precision measurement of time-reversal symmetry violation with laser-cooled
  polyatomic molecules},}\ }\href {\doibase 10.1103/PhysRevLett.119.133002}
  {\bibfield  {journal} {\bibinfo  {journal} {Phys. Rev. Lett.}\ }\textbf
  {\bibinfo {volume} {119}},\ \bibinfo {pages} {133002} (\bibinfo {year}
  {2017})}\BibitemShut {NoStop}%
\bibitem [{\citenamefont {Patterson}\ and\ \citenamefont
  {Doyle}(2007)}]{Patterson:JCP126:154307}%
  \BibitemOpen
  \bibfield  {author} {\bibinfo {author} {\bibfnamefont {David}\ \bibnamefont
  {Patterson}}\ and\ \bibinfo {author} {\bibfnamefont {John~M.}\ \bibnamefont
  {Doyle}},\ }\bibfield  {title} {\enquote {\bibinfo {title} {Bright, guided
  molecular beam with hydrodynamic enhancement},}\ }\href {\doibase
  10.1063/1.2717178} {\bibfield  {journal} {\bibinfo  {journal} {J. Chem.
  Phys.}\ }\textbf {\bibinfo {volume} {126}},\ \bibinfo {pages} {154307}
  (\bibinfo {year} {2007})}\BibitemShut {NoStop}%
\bibitem [{\citenamefont {Lu}\ \emph {et~al.}(2011)\citenamefont {Lu},
  \citenamefont {Rasmussen}, \citenamefont {Wright}, \citenamefont
  {Patterson},\ and\ \citenamefont {Doyle}}]{Lu:PCCP13:18986}%
  \BibitemOpen
  \bibfield  {author} {\bibinfo {author} {\bibfnamefont {Hsin-I}\ \bibnamefont
  {Lu}}, \bibinfo {author} {\bibfnamefont {Julia}\ \bibnamefont {Rasmussen}},
  \bibinfo {author} {\bibfnamefont {Matthew~J.}\ \bibnamefont {Wright}},
  \bibinfo {author} {\bibfnamefont {Dave}\ \bibnamefont {Patterson}}, \ and\
  \bibinfo {author} {\bibfnamefont {John~M.}\ \bibnamefont {Doyle}},\
  }\bibfield  {title} {\enquote {\bibinfo {title} {A cold and slow molecular
  beam},}\ }\href {\doibase 10.1039/C1CP21206K} {\bibfield  {journal} {\bibinfo
   {journal} {Phys. Chem. Chem. Phys.}\ }\textbf {\bibinfo {volume} {13}},\
  \bibinfo {pages} {18986--18990} (\bibinfo {year} {2011})}\BibitemShut
  {NoStop}%
\bibitem [{\citenamefont {Lim}\ \emph {et~al.}(2015)\citenamefont {Lim},
  \citenamefont {Frye}, \citenamefont {Hutson},\ and\ \citenamefont
  {Tarbutt}}]{Lim:PRA92:053419}%
  \BibitemOpen
  \bibfield  {author} {\bibinfo {author} {\bibfnamefont {Jongseok}\
  \bibnamefont {Lim}}, \bibinfo {author} {\bibfnamefont {Matthew~D.}\
  \bibnamefont {Frye}}, \bibinfo {author} {\bibfnamefont {Jeremy~M.}\
  \bibnamefont {Hutson}}, \ and\ \bibinfo {author} {\bibfnamefont {M.~R.}\
  \bibnamefont {Tarbutt}},\ }\bibfield  {title} {\enquote {\bibinfo {title}
  {Modeling sympathetic cooling of molecules by ultracold atoms},}\ }\href
  {\doibase 10.1103/PhysRevA.92.053419} {\bibfield  {journal} {\bibinfo
  {journal} {Phys. Rev. A}\ }\textbf {\bibinfo {volume} {92}},\ \bibinfo
  {pages} {053419} (\bibinfo {year} {2015})}\BibitemShut {NoStop}%
\bibitem [{\citenamefont {Pobell}(2007)}]{Pobell:17675}%
  \BibitemOpen
  \bibfield  {author} {\bibinfo {author} {\bibfnamefont {Frank}\ \bibnamefont
  {Pobell}},\ }\href
  {http://gen.lib.rus.ec/book/index.php?md5=6CE684A6F021FE63F7626F4C0B735BEC}
  {\emph {\bibinfo {title} {Matter and methods at low temperatures}}},\
  \bibinfo {edition} {3rd}\ ed.\ (\bibinfo  {publisher} {Springer},\ \bibinfo
  {year} {2007})\BibitemShut {NoStop}%
\bibitem [{\citenamefont {Friedrich}\ \emph {et~al.}(1998)\citenamefont
  {Friedrich}, \citenamefont {deCarvalho}, \citenamefont {Kim}, \citenamefont
  {Patterson}, \citenamefont {Weinstein},\ and\ \citenamefont
  {Doyle}}]{Bretislav:JCS94:1783}%
  \BibitemOpen
  \bibfield  {author} {\bibinfo {author} {\bibfnamefont {Bretislav}\
  \bibnamefont {Friedrich}}, \bibinfo {author} {\bibfnamefont {Robert}\
  \bibnamefont {deCarvalho}}, \bibinfo {author} {\bibfnamefont {Jinha}\
  \bibnamefont {Kim}}, \bibinfo {author} {\bibfnamefont {David}\ \bibnamefont
  {Patterson}}, \bibinfo {author} {\bibfnamefont {Jonathan~D.}\ \bibnamefont
  {Weinstein}}, \ and\ \bibinfo {author} {\bibfnamefont {John~M.}\ \bibnamefont
  {Doyle}},\ }\bibfield  {title} {\enquote {\bibinfo {title} {Towards magnetic
  trapping of molecules},}\ }\href {\doibase 10.1039/A708859K} {\bibfield
  {journal} {\bibinfo  {journal} {J. Chem. Soc.{,} Faraday Trans.}\ }\textbf
  {\bibinfo {volume} {94}},\ \bibinfo {pages} {1783--1791} (\bibinfo {year}
  {1998})}\BibitemShut {NoStop}%
\bibitem [{\citenamefont {Singh}\ \emph {et~al.}(2018)\citenamefont {Singh},
  \citenamefont {Samanta}, \citenamefont {Roth}, \citenamefont {Gusa},
  \citenamefont {Ossenbr\"uggen}, \citenamefont {Rubinsky}, \citenamefont
  {Horke},\ and\ \citenamefont {K\"upper}}]{Singh:PRA97:032704}%
  \BibitemOpen
  \bibfield  {author} {\bibinfo {author} {\bibfnamefont {Vijay}\ \bibnamefont
  {Singh}}, \bibinfo {author} {\bibfnamefont {Amit~K.}\ \bibnamefont
  {Samanta}}, \bibinfo {author} {\bibfnamefont {Nils}\ \bibnamefont {Roth}},
  \bibinfo {author} {\bibfnamefont {Daniel}\ \bibnamefont {Gusa}}, \bibinfo
  {author} {\bibfnamefont {Tim}\ \bibnamefont {Ossenbr\"uggen}}, \bibinfo
  {author} {\bibfnamefont {Igor}\ \bibnamefont {Rubinsky}}, \bibinfo {author}
  {\bibfnamefont {Daniel~A.}\ \bibnamefont {Horke}}, \ and\ \bibinfo {author}
  {\bibfnamefont {Jochen}\ \bibnamefont {K\"upper}},\ }\bibfield  {title}
  {\enquote {\bibinfo {title} {Optimized cell geometry for buffer-gas-cooled
  molecular-beam sources},}\ }\href {\doibase 10.1103/PhysRevA.97.032704}
  {\bibfield  {journal} {\bibinfo  {journal} {Phys. Rev. A}\ }\textbf {\bibinfo
  {volume} {97}},\ \bibinfo {pages} {032704} (\bibinfo {year}
  {2018})}\BibitemShut {NoStop}%
\bibitem [{\citenamefont {Huang}\ and\ \citenamefont
  {Chen}(2006)}]{HUANG:CR46:833}%
  \BibitemOpen
  \bibfield  {author} {\bibinfo {author} {\bibfnamefont {Y.H.}\ \bibnamefont
  {Huang}}\ and\ \bibinfo {author} {\bibfnamefont {G.B.}\ \bibnamefont
  {Chen}},\ }\bibfield  {title} {\enquote {\bibinfo {title} {A practical vapor
  pressure equation for helium-3 from 0.01 {K} to the critical point},}\ }\href
  {\doibase https://doi.org/10.1016/j.cryogenics.2006.07.006} {\bibfield
  {journal} {\bibinfo  {journal} {Cryogenics}\ }\textbf {\bibinfo {volume}
  {46}},\ \bibinfo {pages} {833 -- 839} (\bibinfo {year} {2006})}\BibitemShut
  {NoStop}%
\bibitem [{Note1()}]{Note1}%
  \BibitemOpen
  \bibinfo {note} {From the relation, $dP/dT = LP/RT^2$, the cooling power of
  an evaporation refrigerator is proportional to the vapor pressure (P). To
  achieve higher cooling power at low temperatures, a $^3$He evaporation
  refrigerator is preferred over that $^4$He. Further, the evaporation
  refrigerators are operated in closed cycles and hence there is no $^3$He
  consumption. This brings down the operational cost of a $^3$He refrigerator
  similar to that of a $^4$He refrigerator. Currently, $^4$He evaporation
  refrigerators are operating at the Harvard University and the University of
  the Nevada Reno (UNR). The technical challenges of developing such
  refrigerators have already been overcome~\cite
  {Singh:PRL108:203201}.}\BibitemShut {Stop}%
\bibitem [{Note2()}]{Note2}%
  \BibitemOpen
  \bibinfo {note} {The second stage of the cryogenic refrigerator
  (CRYOMECH:PT420) has a cooling power of 2~W at 4.2~K
  (http://www.cryomech.com/cryorefrigerators/pulse-tube/). As shown in
  the~\protect \autoref {sec:appendix}, the total heat load on the second stage
  of the cryogenic refrigerator is much smaller than this cooling
  power.}\BibitemShut {Stop}%
\bibitem [{Note3()}]{Note3}%
  \BibitemOpen
  \bibinfo {note} {As shown in the~\protect \autoref {sec:appendix}, the
  calculated heat load on the $^3$He cryogenic refrigerator stays below 1~mW,
  when the beam source is operated with a repetition rate of 20~Hz}\BibitemShut
  {NoStop}%
\bibitem [{\citenamefont {Piskorski}\ \emph {et~al.}(2014)\citenamefont
  {Piskorski}, \citenamefont {Patterson}, \citenamefont {Eibenberger},\ and\
  \citenamefont {Doyle}}]{Piskorski:CPC15:3800}%
  \BibitemOpen
  \bibfield  {author} {\bibinfo {author} {\bibfnamefont {Julia}\ \bibnamefont
  {Piskorski}}, \bibinfo {author} {\bibfnamefont {David}\ \bibnamefont
  {Patterson}}, \bibinfo {author} {\bibfnamefont {Sandra}\ \bibnamefont
  {Eibenberger}}, \ and\ \bibinfo {author} {\bibfnamefont {John~M.}\
  \bibnamefont {Doyle}},\ }\bibfield  {title} {\enquote {\bibinfo {title}
  {Cooling, spectroscopy and non‐sticking of trans‐stilbene and nile
  red},}\ }\href {\doibase 10.1002/cphc.201402502} {\bibfield  {journal}
  {\bibinfo  {journal} {ChemPhysChem}\ }\textbf {\bibinfo {volume} {15}},\
  \bibinfo {pages} {3800--3804} (\bibinfo {year} {2014})}\BibitemShut {NoStop}%
\bibitem [{\citenamefont {Maussang}\ \emph {et~al.}(2005)\citenamefont
  {Maussang}, \citenamefont {Egorov}, \citenamefont {Helton}, \citenamefont
  {Nguyen},\ and\ \citenamefont {Doyle}}]{Maussang:PRL94:123002}%
  \BibitemOpen
  \bibfield  {author} {\bibinfo {author} {\bibfnamefont {Kenneth}\ \bibnamefont
  {Maussang}}, \bibinfo {author} {\bibfnamefont {Dima}\ \bibnamefont {Egorov}},
  \bibinfo {author} {\bibfnamefont {Joel~S.}\ \bibnamefont {Helton}}, \bibinfo
  {author} {\bibfnamefont {Scott~V.}\ \bibnamefont {Nguyen}}, \ and\ \bibinfo
  {author} {\bibfnamefont {John~M.}\ \bibnamefont {Doyle}},\ }\bibfield
  {title} {\enquote {\bibinfo {title} {Zeeman relaxation of {CaF} in
  low-temperature collisions with helium},}\ }\href {\doibase
  10.1103/PhysRevLett.94.123002} {\bibfield  {journal} {\bibinfo  {journal}
  {Phys. Rev. Lett.}\ }\textbf {\bibinfo {volume} {94}},\ \bibinfo {pages}
  {123002} (\bibinfo {year} {2005})}\BibitemShut {NoStop}%
\bibitem [{\citenamefont {Hutzler}\ \emph {et~al.}(2011)\citenamefont
  {Hutzler}, \citenamefont {Parsons}, \citenamefont {Gurevich}, \citenamefont
  {Hess}, \citenamefont {Petrik}, \citenamefont {Spaun}, \citenamefont {Vutha},
  \citenamefont {DeMille}, \citenamefont {Gabrielse},\ and\ \citenamefont
  {Doyle}}]{Hutzler:PCCP13:18976}%
  \BibitemOpen
  \bibfield  {author} {\bibinfo {author} {\bibfnamefont {Nicholas~R.}\
  \bibnamefont {Hutzler}}, \bibinfo {author} {\bibfnamefont {Maxwell~F.}\
  \bibnamefont {Parsons}}, \bibinfo {author} {\bibfnamefont {Yulia~V.}\
  \bibnamefont {Gurevich}}, \bibinfo {author} {\bibfnamefont {Paul~W.}\
  \bibnamefont {Hess}}, \bibinfo {author} {\bibfnamefont {Elizabeth}\
  \bibnamefont {Petrik}}, \bibinfo {author} {\bibfnamefont {Ben}\ \bibnamefont
  {Spaun}}, \bibinfo {author} {\bibfnamefont {Amar~C.}\ \bibnamefont {Vutha}},
  \bibinfo {author} {\bibfnamefont {David}\ \bibnamefont {DeMille}}, \bibinfo
  {author} {\bibfnamefont {Gerald}\ \bibnamefont {Gabrielse}}, \ and\ \bibinfo
  {author} {\bibfnamefont {John~M.}\ \bibnamefont {Doyle}},\ }\bibfield
  {title} {\enquote {\bibinfo {title} {A cryogenic beam of refractory{,}
  chemically reactive molecules with expansion cooling},}\ }\href {\doibase
  10.1039/C1CP20901A} {\bibfield  {journal} {\bibinfo  {journal} {Phys. Chem.
  Chem. Phys.}\ }\textbf {\bibinfo {volume} {13}},\ \bibinfo {pages}
  {18976--18985} (\bibinfo {year} {2011})}\BibitemShut {NoStop}%
\bibitem [{Note4()}]{Note4}%
  \BibitemOpen
  \bibinfo {note} {The time taken by the beam in traveling 1~cm is 0.09~ms. The
  transverse distance traveled by the beam during this time is $\pm $ 0.2
  mm.}\BibitemShut {Stop}%
\bibitem [{Com()}]{Comsol:Multiphysics:5.3}%
  \BibitemOpen
  \href@noop {} {}\bibinfo {note} {COMSOL Multiphysics v.\ 5.3.\
  \url{http://www.comsol.com}. COMSOL AB, Stockholm, Sweden}\BibitemShut
  {NoStop}%
\bibitem [{Note5()}]{Note5}%
  \BibitemOpen
  \bibinfo {note} {This heat load is calculated using the literature value of
  the CaF$_2$ specific heat of
  854~J~Kg$^{-1}$~K$^{-1}$~(https://www.crystran.co.uk/optical-materials/calcium-
  fluoride-caf2)}\BibitemShut {NoStop}%
\end{thebibliography}%

\section{Appendix}
\label{sec:appendix}
This section presents the calculations of the Maxwell-Boltzmann PDF of the speeds of the helium atoms at low temperatures, simulation of helium behavior inside the buffer gas cells, and the required helium flow rate for 10~mW cooling power of the helium bath. The Maxwell-Boltzmann PDF of helium atoms indicates that the fraction of the molecules moving with speeds below 10~m/s can be significantly enhanced by lowering the cell temperature. Furthermore, the simulated helium behavior inside the buffer gas cells shows that the diffusion losses to the cold cell walls can be minimized by using the buffer gas cells with conical geometry. On the other hand, the calculated heat loads on the first and second stages of the cryogenic refrigerator as well as on the $^3$He bath are well below their corresponding cooling powers. Thus, it is technically feasible to design a two-stage buffer gas-cooled beam source with the parameters considered here.

\subsection{Maxwell-Boltzmann probability density function of helium atoms at low temperatures}
\label{sec:PDF}
\begin{figure}
   \centering%
   \includegraphics[width=\linewidth]{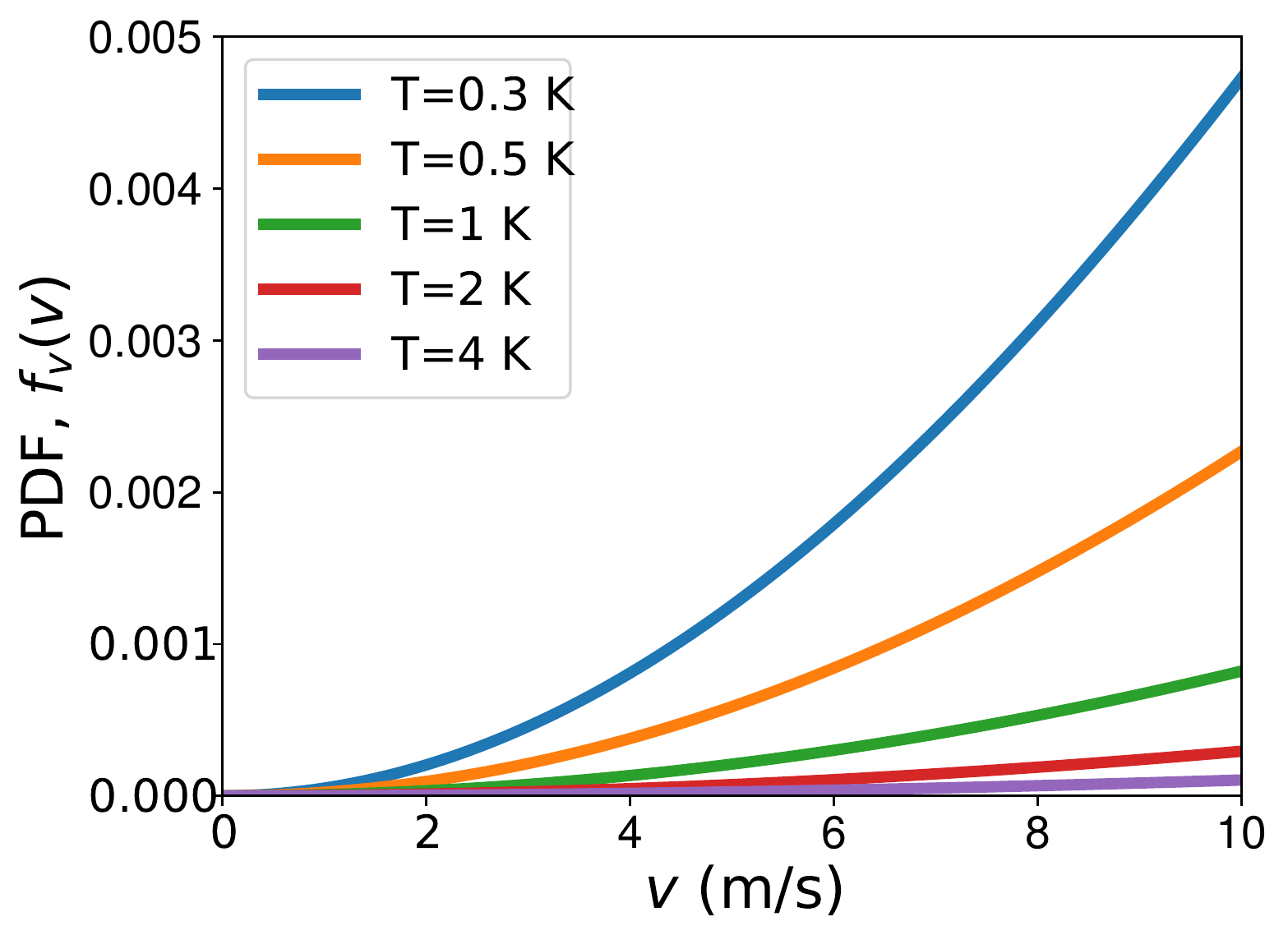}
   \caption{Calculated Maxwell-Boltzmann PDF of helium buffer gas atoms at low temperatures. The fraction of the slower atoms increases rapidly with decrease in the buffer gas temperature.}
   \label{fig:maxwell}
\end{figure}

For helium buffer gas atoms, we have calculated the Maxwell-Boltzmann PDF, f$_\text{v}(v)$, using the following equation. 
\begin{equation}
f_\text{v}(v) = 4\pi v^2 \sqrt{\frac{m}{2\pi k_\text{B}T}}\exp(-\frac{mv^2}{2k_\text{B}T})
\end{equation}
The calculated PDF for helium atoms moving with a speed below 10~m/s at low temperatures is shown in~\autoref{fig:maxwell}. As expected, the fraction of the slower molecules increases rapidly with decrease in temperature.   
The calculated sums of the PDFs for 0 - 5~m/s, 0 - 10~m/s, 0 - 20~m/s, and 0 - 50~m/s speed ranges at low temperatures ranging from 0.3~K to 4~K are shown in~\autoref{tab:max}. In the hydrodynamic entrainment regime, the speed distribution of the buffer gas cooled molecules is similar to that of buffer gas atoms. Thus, these PDF values are used for calculating the flux of the slower molecules in the molecular beams produced from our two-stage buffer gas cooled beam source. The total flux of the CaF produced from this source is 8~$\times$~10$^{12}$ molecules per pulse. Using this flux and PDF values of helium at 0.5~K, the flux of the CaF moving with speeds below 5~m/s, 10~m/s, 20~m/s, and 50~m/s are calculated to be 8~$\times$~10$^{9}$, 6~$\times$~10$^{10}$, 4~$\times$~10$^{11}$, and 4~$\times$~10$^{12}$ molecules per pulse respectively.     
\begin{table*}
   \caption{Calculated f$_\text{v}(v)$ for helium atoms from 0.3~K to 4~K.}
   \begin{tabular}{ccccc}
     \hline\hline
     T~(K) & f$_\text{v}$(0-5~m/s) & f$_\text{v}$(0-10~m/s) & f$_\text{v}$(0-20~m/s) & f$_\text{v}$(0-50~m/s)\\
     \hline
     0.3 & 0.002 & 0.02 & 0.11 & 0.74\\
     0.5 & 0.001 & 0.008 & 0.06 & 0.51\\
     1 & 0.0004 & 0.003 & 0.02 & 0.3\\
     2 & 0.0001 & 0.001 & 0.008 & 0.1\\
     4 & 0.00004 & 0.0004 & 0.003 & 0.04\\
     \hline
   \end{tabular} 
   \label{tab:max}
\end{table*}
\subsection{Simulation of helium behavior inside the buffer gas cells}
\label{sec:Sim}
\begin{figure}
   \centering%
   \includegraphics[width=\linewidth]{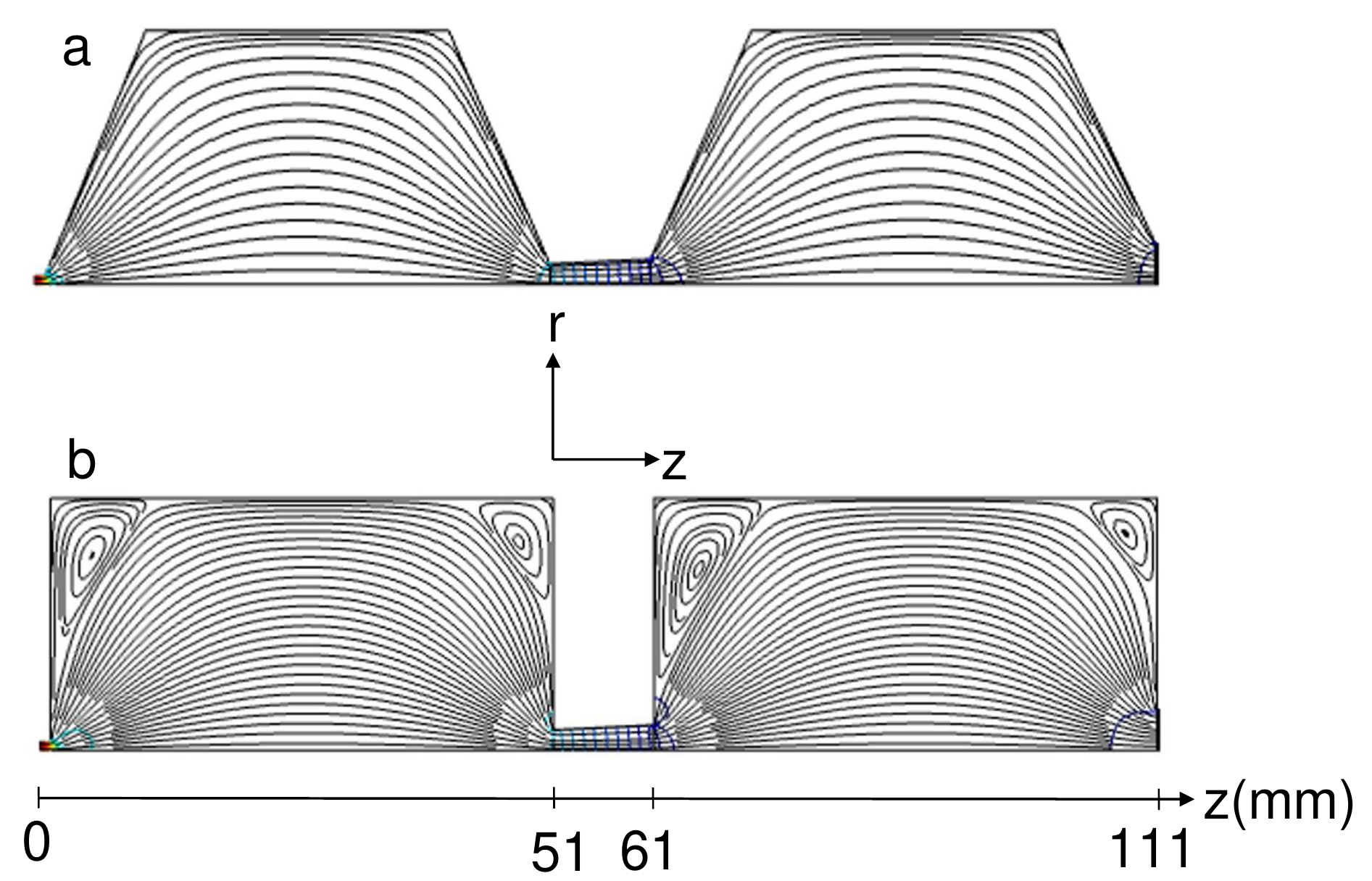}
   \caption{Simulations of the helium flow-field for two different geometries of the cryogenic cells. (a) Cells with conical entrance 
   and conical exit. (b) Cells with planar entrance and planar exit. Depicted are streamlines with the helium flowing from left to 
   right. For clarity, only one half of the cylindrically-symmetric r, z-plane is shown.}
   \label{fig:sim}
\end{figure}

To determine the optimum geometry of our buffer gas cells, we have simulated the helium flow-fields inside the buffer gas cells using a finite-element solver, COMSOL Multiphysics with laminar flow interface~\cite{Comsol:Multiphysics:5.3}, treating helium as an ideal gas in the temperature range 3-0.5 K. For 12~SCCM of helium flow and a pressure of $10^{-6}$~mbar outside the cell exit, the resulting flow fields are shown in~\autoref{fig:sim} as slices through the cylindrically-symmetric cell volume, i.e., in $r,z$ space.
From the simulations, we have extracted the helium pressures inside the cells. These extracted values are shown in~\autoref{tab:sim} and are in a good qualitative agreement with the calculations. 
\begin{table*}
   \caption{Simulated helium pressures at 12~SCCM helium flow rate.}
   \begin{tabular}{ccc}
     \hline\hline
     region & pressure~(mbar)& density~(cm$^{-3}$)\\
     \hline
     3~K cell ($z$ = 1 to $z$ = 51) &0.03 &7.2$\times$10$^{16}$\\
     between the cells ($z$ = 51 to $z$ = 61) &0.02 &4.8$\times$10$^{16}$\\
     0.5~K cell ($z$ = 61 to $z$ = 111) &0.003 &4.3$\times$10$^{16}$\\
     0.5~K cell exit ($z$ = 111) &0.002 &2.8$\times$10$^{16}$\\
     \hline
   \end{tabular}
   \label{tab:sim}
\end{table*}

Furthermore, the simulations show that the planar cell geometries produce vortices in the flow fields 
at the cell corners. This can be circumvented by the use of conical cell geometry, which results in an 
overall laminar flow. The hot CaF molecules introduced into the cell are thermalized with the buffer-gas 
before reaching to cell exit and hence follow the streamlines. The presence of  vortices in the flow-field 
can therefore lead to trapping of the molecules and eventually loss to the cell walls. On the other hand, 
the laminar flow produced by a conical cell geometry avoids these diffusion losses and, therefore, allows 
entraining maximum number of the molecules into the beam~\citep{Singh:PRA97:032704}. Thus, the buffer gas 
cells with conical extrance and conical exit are beneficial for extracting the maximum number of the molecules 
from the proposed two-stage buffer gas-cooled beam source.
\subsection{Calculated mass flow rate for 10 mW cooling power of the helium bath}
At 0.5~K, the $^3$He vapor pressure = 0.2~mbar~\cite{Pobell:17675,HUANG:CR46:833}. The latent 
heat of vaporization of the $^3$He is $L_\text{v}$ = 8.5~J/gm at 3.2~K temperature 
and 1 bar pressure. One gram of $^3$He contains 2 $\times$ 10$^{23}$ atoms. 

Thus, the latent heat of vaporization $L_\text{v}$ = 4.25 $\times$ 10$^{-23}$~J/atom. 
The number of atoms ($N_\text{A}$) contained by 1~L of $^3$He at $T$ = 0.5~K and 
$P$ = 0.2~mbar can be calculated from the ideal gas law $PV = N_\text{A} k_{B} T$. 
$N_\text{A}$ = (0.2~mbar)(1~L)/((1.38$\times$ 10$^{-23}$~J/K)(0.5~K)) = 
(20~Pa)(0.001~m$^3$)/((1.38 $\times$ 10$^{-23}$~J/K)(0.5~K)) = 30~$\times$ 10$^{20}$. 
Thus, at $T$ = 0.5~K and $P$ = 0.2~mbar, $L_\text{v}$ = (4.25 $\times$ 10$^{-23}$~J/atom)
(30 $\times$ 10$^{20}$~atom/L) = 0.13~J/L. 

To achieve a cooling power of 10~mW, the boil off rate = (10~mW/0.13~J/L) = 0.078~L/s. 
Operating the refrigerator continuously, the liquid $^3$He flow into the helium bath 
must be equal to the boil off rate. Thus, the required flow through the helium condenser 
(3~K, 1 bar) calculated using ideal gas law = (0.2~mbar/1~bar) (3~K/0.5~K) (0.078~L/s)= 
9.4 $\times$ 10$^{-5}$~L/s. At $T$ = 3~K and $P$ = 1 bar, this corresponds to a gas flow 
of $N_\text{A} = PV/k_\text{B}T$ = (100000 Pa)(9.4 $\times$ 10$^{-5}$~L/s)(0.001 m$^3$/L)/ ((1.38 $
\times$ 10$^{-23}$~J/K) (3~K)) = 2.3 $\times$ 10$^{20}$ atom/s = (2.3 $\times$ 10$^{20}$ atom/s) 
(1~gm/2 $\times$ 10$^{23}$ atom) = 1.2~mg/s.

\subsection{Calculated heat load on the first stage of the refrigerator}
At the first stage both $^3$He (evaporated from the helium bath) and $^4$He (used for helium 
buffer gas cooling) are required to be cooled down from 300~K to 30~K. The $^3$He mass flow 
rate = 1.2~mg/s. The $^4$He flow rate = 20~SCCM (over estimated) = (20 $\times$ 4.5 $\times$ 
10$^{17}$)(4~gm/s)/(6.023 $\times$ 10$^{23}$) = 0.06~mg/s. The total heat load on the first stage 
of the cryogenic refrigerator = ((1.2 + 0.06)~mg/s)(5.2~J/(gm K))(270~K) = 1.8~W. This load is 
much smaller than the cooling power (55~W at 45~K) of the cryogenic refrigerator (CRYOMECH: PT420).

\subsection{Calculated heat load on the second stage of the refrigerator}
At the second stage, the $^3$He required to cool from 30 to 3~K and convert from gas to liquid, 
the $^4$He needs to be cooled from 30 to 3~K, and molecules are vaporized by 10~mJ laser pulse 
with 20~Hz repetition rate by laser ablation in the first stage of the buffer gas cell attached 
to the second stage. The the total heat load on the second stage = (1.2~mg/s)(5.2~J/(gm K))(27~K) + 
(1.2~mg/s)(8.5~J/gm) + (0.06~mg/s)(5.2~J/(gm K))(27~K) + (10~mJ)(20~pulse/s) = 0.4~W. This heat load 
is significantly smaller compared to the second stage cooling power (2~W at 4.2~K) of the cryogenic 
refrigerator.

\subsection{Calculated heat load on the 0.5~K stage of the buffer gas cell}
The 20~SCCM flow from the first stage of the buffer gas cell required to be cooled down from 3~K to 
0.5~K on the second stage of the buffer gas cell attached to the helium bath. Thus, the heat load on 
the 0.5~K helium bath = (0.06~mg/s)(5.2~J/(gm K))(2.5~K) = 0.8~mW. Further, the heat load from 
cooling of the 2 $\times$ 10$^{13}$ CaF molecules from 3~K to 0.5~K with 20~Hz repetition rate is 
below 0.1~mW~\footnote{This heat load is calculated using the literature value of the CaF$_2$ 
specific heat of 854~J~Kg$^{-1}$~K$^{-1}$~(https://www.crystran.co.uk/optical-materials/calcium-
fluoride-caf2)}. Thus, the total heat load on the 0.5~K stage is much smaller compared to the 
targeted cooling power (10~mW) of the helium bath.
\end{document}